\documentclass[aps, pra, reprint, superscriptaddress]{revtex4-1}

\usepackage{amsmath}
\usepackage{amssymb}
\usepackage{wasysym}
\usepackage{graphicx}
\usepackage{hyperref}
\usepackage{color}
\usepackage{physics}
\usepackage{siunitx}
\usepackage{xcolor}
\usepackage{changes}
\usepackage{comment}
\usepackage{siunitx}
\usepackage{bm}

\allowdisplaybreaks

\begin{document}

\title{Long-lived fermionic Feshbach molecules with tunable $p$-wave interactions}

\author{Marcel~Duda}
\thanks{These two authors contributed equally.}
\author{Xing-Yan~Chen}
\thanks{These two authors contributed equally.}
\author{Roman~Bause}
\author{Andreas~Schindewolf}
\affiliation{Max-Planck-Institut f\"{u}r Quantenoptik, 85748 Garching, Germany}
\affiliation{Munich Center for Quantum Science and Technology, 80799 M\"{u}nchen, Germany}
\author{Immanuel~Bloch}
\affiliation{Max-Planck-Institut f\"{u}r Quantenoptik, 85748 Garching, Germany}
\affiliation{Munich Center for Quantum Science and Technology, 80799 M\"{u}nchen, Germany}
\affiliation{Fakult\"{a}t f\"{u}r Physik, Ludwig-Maximilians-Universit\"{a}t, 80799 M\"{u}nchen, Germany}
\author{Xin-Yu~Luo}
\email[]{E-Mail: xinyu.luo@mpq.mpg.de}
\affiliation{Max-Planck-Institut f\"{u}r Quantenoptik, 85748 Garching, Germany}
\affiliation{Munich Center for Quantum Science and Technology, 80799 M\"{u}nchen, Germany}

\date{\today}

\begin{abstract}
Ultracold fermionic Feshbach molecules are promising candidates for exploring quantum matter with strong $p$-wave interactions, however, their lifetimes were measured to be short. Here, we characterize the $p$-wave collisions of ultracold fermionic $^{23}\mathrm{Na}^{40}\mathrm{K}$ Feshbach molecules for different scattering lengths and temperatures. By increasing the binding energy of the molecules, the two-body loss coefficient reduces by three orders of magnitude leading to a second-long lifetime, 20 times longer than that of ground-state molecules. We exploit the scaling of elastic and inelastic collisions with the scattering length and temperature to identify a regime where the elastic collisions dominate over the inelastic ones allowing the molecular sample to thermalize. Our work provides a benchmark for four-body calculations of molecular collisions and is essential for producing a degenerate Fermi gas of Feshbach molecules.
\end{abstract}

\maketitle

Ultracold molecules have gained considerable attention in the past years. Their rich internal structure offers unique opportunities for quantum-engineering and quantum-chemistry applications \cite{Carr2009,Quemener2012,Bohn2017}. However, a major challenge for their usability to investigate quantum many-body physics is to identify interacting molecular systems where the undesired collisional loss processes are under control.

A promising platform are optically trapped Feshbach molecules which are formed through an atomic interspecies Feshbach resonance \cite{Chin2010}. These molecules do not only represent a key intermediate product for generating ground-state polar molecules, but also are intriguing due to their rich collisional behavior near the Feshbach resonance. When the molecules are weakly bound and the wavefunction extends beyond the interatomic potential, the only relevant length scale in the system is the interspecies scattering length. It determines the size and the interactions between these so-called universal halo dimers \cite{Ferlaino2008}. Molecules composed of one boson and one fermion are especially interesting as the dominating collision channel in these composite fermions is $p$-wave. As a consequence, the molecules are protected by a centrifugal $p$-wave barrier from reaching short range and undergoing inelastic loss, which has been confirmed for ground-state molecules \cite{Ospelkaus2010, Bause2021b}. In addition, the elastic $p$-wave collisions are tunable with the interspecies scattering length and with temperature. These scalings are expected to be stronger than for the inelastic collisions \cite{Marcelis2008}. Therefore, fermionic Feshbach molecules have been proposed to exhibit unconventional superfluidity \cite{Kenji2009,Bazak2018}.

Common wisdom dictates that Feshbach molecules, being in a highly excited vibrational state, should be short-lived, especially when compared to molecules in the rovibronic ground state. In addition, unlike bosonic molecules composed of fermions \cite{Greiner2003,Regal2004}, Pauli blocking does not protect fermionic Feshbach molecules from undergoing three-body loss, involving two bosons and one fermion colliding at short range. The collisional instability of fermionic Feshbach molecules was confirmed in experiments where lifetimes on the order of hundred milliseconds \cite{Zirbel2008,Wu2012} or as short as a millisecond \cite{Heo2012} were measured. Due to the short lifetime of these molecules, so far no experimental evidence for elastic $p$-wave collisions could be provided despite the favorable scaling of elastic-to-inelastic collisions with temperature and scattering length.

In this work, we study the $p$-wave inelastic and elastic collisions in a trapped sample of fermionic $\mathrm{NaK}$ Feshbach molecules. We tune the interspecies scattering length by changing the magnetic field field near a Feshbach resonance allowing us to change the inelastic collision rate by three orders of magnitude. In addition, we observe a linear dependence of the collisional loss with temperature. We study the elastic collisions by measuring the cross-dimensional thermalization in out-of-equilibrium molecular samples. In particular, we can exploit the stronger scaling of elastic collisions compared to the inelastic collisions on the scattering length and temperature to identify a regime where elastic collisions dominate.

In a first set of measurements, we characterize the loss between NaK Feshbach molecules which occurs when the molecules overcome the centrifugal $p$-wave barrier. Once they reach the short-range regime, the loss can be described by Na-Na-K three-body recombination where the additional K atom acts as a spectator and carries away the kinetic energy on the order of the Feshbach-molecule binding energy \cite{privateDima2020}. 

We begin by preparing a pure sample of Feshbach molecules in a crossed optical dipole trap. We start from a mixture of bosonic $\mathrm{^{23}Na}$ and fermionic $\mathrm{^{40}K}$ atoms in their respective energetically lowest hyperfine states $|F,m_F \rangle = |1,1 \rangle$ and $|9/2,-9/2 \rangle$, where $F$ represents the total angular momentum and $m_F$ the projection onto the magnetic-field axis. To associate Feshbach molecules, we ramp the magnetic field across an interspecies Feshbach resonance at \SI{78.3}{G}, which we have previously characterized \cite{Chen2021}, followed by an additional fast magnetic-field ramp to \SI{72.3}{G} where the Feshbach molecules have a vanishing magnetic moment. As the remaining atoms possess a finite magnetic moment, we remove them from the trap by applying a magnetic-field gradient of \SI{40}{G/cm}. This procedure typically produces $3 \times 10^{4}$ Feshbach molecules at a temperature of \SI{500}{nK}. 

We investigate the dependence of the collisional loss on the scattering length  by ramping the magnetic field to the target magnetic field $B$  where we hold the molecules for a variable time. For detection, we first turn off all dipole traps, then ramp the magnetic field back through the Feshbach resonance to dissociate the molecules. Finally, the dissociated atoms are detected in time of flight using absorption imaging.

We fit the loss of Feshbach molecules in a harmonic trap with the two-body loss model
\begin{align}
\label{eq:density-loss}
\frac{\mathrm{d}N}{\mathrm{dt}} = -\frac{\beta_{\mathrm{inel}} N^2}{\nu_0 T^{3/2}} \, ,
\end{align}
where $\beta_{\mathrm{inel}}$ is the inelastic collision coefficient, $N$ is the number of molcules, $\nu_0T^{3/2}$ is the generalized trap volume given by $\nu_0 = (4\pi k_B / m\bar{\omega}^2)^{3/2}$, with $\bar{\omega} = (\omega_x\omega_y\omega_z)^{1/3}$ the geometric mean of the trap frequencies, $k_B$ the Boltzmann constant and $m$ the mass of molecules. Since we do not observe a significant change of the temperature during the measurement, we do not include any heating effects in our model. We also ignore the inelastic collisions between Feshbach molecules and spectator K atoms that remain in the trap close to the resonance where their collisions are strongly suppressed by Pauli blocking \cite{Zirbel2008}.

We extract the interspecies scattering length from the magnetic field using a model with two overlapping Feshbach resonances given by 
\begin{align}
\label{eq:Btoa}
a(B) = a_{bg}\left( \frac{B-B^{*}_{1}}{B-B_{0,1}}\right) \left(\frac{B-B^{*}_{2}}{B-B_{0,2}}\right) \, ,
\end{align}
where $a_{bg}= - 619\,a_0$ \cite{Viel2016} is the background scattering length, $B_{0,1} = 78.3\,\mathrm{G}$ and $B_{0,2} = 89.7\,\mathrm{G}$ are the two resonance positions, and $B^{*}_{1}=73.03\,\mathrm{G}$ and $B^{*}_{2}=80.36\,\mathrm{G}$ are the zero-crossings of the scattering length \cite{Chen2021}. 

\begin{figure}
\centering
\includegraphics{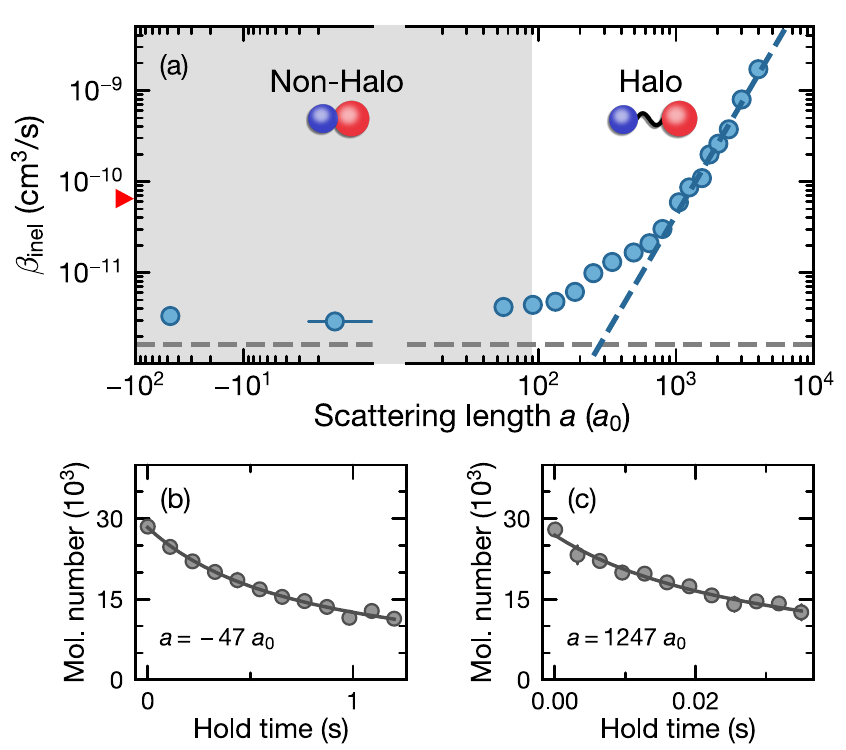}
\caption{Two-body loss coefficient $\beta_{\mathrm{inel}}$ as a function of the scattering length $a$ for $T=\SI{500}{nK}$. (a) Deep in the non-halo regime the loss approaches $\beta_{\mathrm{inel}} = 3.3(1) \times 10^{-12} \,\mathrm{cm^3/s}$.  The gray dashed line indicates the loss coefficient of $\beta_{\mathrm{inel}} = 1.65 \times 10^{-12}\, \mathrm{cm^3/s}$ predicted by MQDT calculations. For $a>1000\,a_0$, $\beta_{\mathrm{inel}}$ is fitted with  
$\beta_{\mathrm{inel}}(a)= c a^l$ which yields $l=2.58(14)$ (blue dashed line). The red triangle indicates the coefficient of inelastic collisions in $\mathrm{NaK}$ ground-state molecules for $T=\SI{500}{nK}$ \cite{Bause2021}. The error in $\beta_{\mathrm{inel}}$ is given by the error of the fit. The horizontal error bars result from a 15-mG uncertainty in the magnetic field. A systematic error of 30\% in $\beta_{\mathrm{inel}}$ is expected from a 10\% error in the trapping frequency.  (b),(c) The number of Feshbach molecules as a function of the hold time for $B= 72\,\mathrm{G}$ ($a=-47\,a_0$) and $B= 77.8\,\mathrm{G}$ ($a=1300\,a_0$). The solid line shows the fitted molecule number from the from the two-body loss model described by Eq.~\eqref{eq:density-loss} assuming a constant temperature.}
\label{fig:loss_a}
\end{figure}

Our findings of $\beta_{\mathrm{inel}}$ as a function of $a$ are summarized in Fig.~\ref{fig:loss_a}. In the halo regime, we observe a strong dependence of the loss coefficient $\beta_{\mathrm{inel}}$ on the scattering length. Calculations for mass-imbalanced systems predict that the inelastic collisional loss should scale as $a^3$ \cite{Marcelis2008}. We check this prediction by fitting the two-body loss coefficient with a polynomial of the form $\beta_{\mathrm{inel}}(a)=c a^l$ for scattering lengths $a > 1000\,a_0$ and obtain $l = 2.58(14)$. All measured rates of inelastic collisions are below the unitarity limit $\beta_{\mathrm{unitary}}= 2 \hbar \lambda_{dB} / \mu $ \cite{Yan2020}, which corresponds to a value of $\beta = 1.77 \times 10^{-9}\,\mathrm{cm^3/s}$ at a temperature of \SI{500}{nK}. Here, $\lambda_{dB}$ corresponds to the de Broglie wavelength a molecule and $\mu$ to the reduced mass of the molecule pair.  

For $a < 1000\,a_0$, the effect of the scattering length on the loss rate is weaker. In particular, once the molecules are deep in the non-halo regime where $a$ is smaller than the van der Waals length $a_{\mathrm{vdW}}$, the two-body loss approaches $\beta_{\mathrm{inel}} = 3.3(1) \times 10^{-12} \,\mathrm{cm^3/s}$ (see Fig.~\ref{fig:loss_a}b). We compare this value to the predictions from multichannel quantum defect theory (MQDT) which have previously reproduced the loss coefficient for molecules in rovibronic ground states \cite{Ospelkaus2010,Bause2021} and in high-lying vibrational states \cite{Hudson2008}. The predicted two-body loss coefficient in collisions of indistinguishable fermions is given by
\begin{align}
\label{eq:MQDT}
\beta_\text{inel}(T) = \frac{\Gamma(1/4)^6}{\Gamma(3/4)^2} \bar{a}^3 \frac{k_B T}{h} = 1513  \bar{a}^3 \frac{k_B T}{h}, \,
\end{align}
where $\bar{a} = 2 \pi \left(2 \mu C_6/\hbar^2\right)^{1/4} / \Gamma(\frac14)^2 = 88.8\,a_0$ is the average scattering length, $C_6$ is the van der Waals coefficient and $\Gamma(x)$ is the gamma function \cite{Idziaszek2010}. The average scattering length $\bar{a}=0.956\,a_{\mathrm{vdW}}$ can be understood as the size of the molecule in the non-halo regime. We calculate the van der Waals coefficient of $\mathrm{NaK}$ Feshbach molecules by approximating the long-range interaction to be the sum of the contribution from individual atoms, i.e.,  $C_{6,\mathrm{FB}} = C_{6,\mathrm{Na}}+C_{6,\mathrm{K}}+2C_{6,\mathrm{Na-K}}$ where $C_{6,\mathrm{Na}} = 1556$, $C_{6,\mathrm{K}} = 3897$ \cite{Derevianko2001} and $C_{6,\mathrm{Na-K}} =2454$ \cite{Mitroy2003} in atomic units. At a temperature of \SI{500}{nK}, Eq.~\ref{eq:MQDT} then yields $1.65 \times 10^{-12}\, \mathrm{cm^3/s}$, which is a factor two lower than the measured value. 

We note that the two-body loss coefficient of $\mathrm{NaK}$ Feshbach molecules in the non-halo regime is 20 times lower than that of the ground-state molecules \cite{Bause2021}. This contradicts the naive expectation that the rovibronic ground-state molecules should live longer than the weakly bound Feshbach molecules. However, this can be explained by considering that the strong molecule-frame dipole moment $d$ present in ground-state molecules modifies the van der Waals interaction, specifically, $C_6$ increases with $d^4$. This effectively increases the van der Waals length for ground-state molecules and thus the universal two-body loss coefficient \cite{Julienne2011}. The agreement of the loss coefficients for ground-state and Feshbach molecules with these models strongly suggests that the lifetime of fermionic bialkali molecules is determined by the long-range interaction potential regardless of the loss mechanism at short range.

\begin{figure}
\centering
\includegraphics{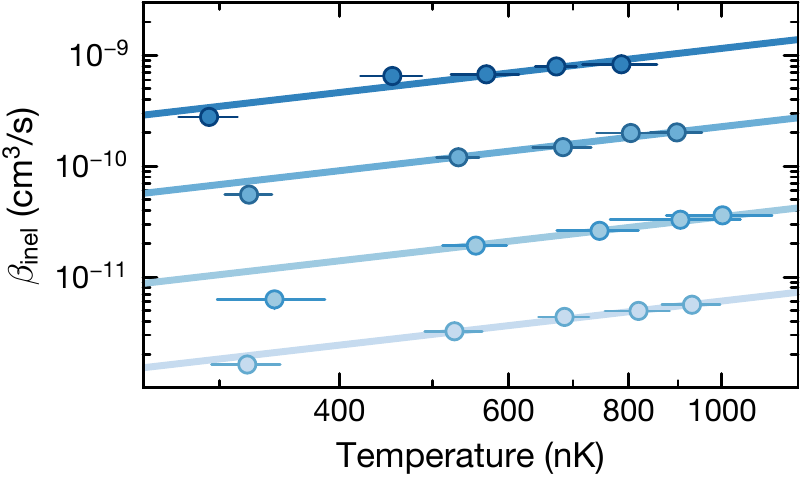}
\caption{Temperature dependence of collisional loss coefficient $\beta_{\mathrm{inel}}$ at $a = -14\,a_0$, $ 554\,a_0$,  $1542\,a_0$, $3000\,a_0$ (from bright to dark blue). The loss coefficient obtained from the respective data sets are fitted with a linear fit (solid lines). The error in $\beta_{\mathrm{inel}}$ is given by the error of the fit. The horizontal error bars represent the statistical error of of $T_\text{avg}$ during the hold time. A systematic error of 30\% in $\beta_{\mathrm{inel}}$ is expected from a 10\% error in the trapping frequency.}
\label{fig:loss_T}
\end{figure}

Next, we study the temperature dependence of the collisional loss. We achieve different temperatures between \SI{300}{nK} and \SI{1000}{nK} by adiabatically compressing or decompressing the trap before the loss measurement. As shown in Fig.~\ref{fig:loss_T}, $\beta_{\mathrm{inel}}$ increases with temperature for all data sets. We fit a linear function to the loss coefficient where we set the $y$-intercept to zero and that the loss coefficient is well described with a linear temperature scaling for all measured scattering lengths. 

In a second set of measurements, we investigate elastic collisions between Feshbach molecules which are predicted to scale as $\beta_{\mathrm{el}} \propto a^6T^{5/2}$ in the halo regime \cite{Marcelis2008}. Thus, we expect that elastic collisions dominate over inelastic collisions for large temperatures and scattering lengths, such that a molecular sample out-of-equilibrium can rethermalize during its lifetime. We create such an out-of-equilibrium molecular sample by non-adiabatically compressing the optical dipole trap after the molecule association such that the trapping frequencies $\omega_y, \omega_z$ increase more than $\omega_x$. After compressing the trap at \SI{77}{G}, we ramp the magnetic field to the target magnetic field $B$ within \SI{1.5}{ms} where we hold the molecules for a variable time. We typically produce $1 \times 10^{4}$ Feshbach molecules with an initial temperature of $T_x = \SI{350}{nK}$ and $T_z=T_y= \SI{550}{nK}$.

To obtain both the elastic and inelastic collision coefficients, we adapt the model used for two-body collisions in polar molecules \cite{Ni2010} to obtain the following set of differential equations:
\begin{align}
&\frac{\mathrm{d}n}{\mathrm{dt}} = -\frac{n^2}{3} K_{\mathrm{inel}} (2T_z+T_x)  - \frac{n}{2 T_x} \frac{\mathrm{d}T_x}{\mathrm{dt}} - \frac{n}{2 T_z} \frac{\mathrm{d}T_z}{\mathrm{dt}}  \, , \label{eqn:ode_elastic_inelastic1} \\
&\frac{\mathrm{d}T_z}{\mathrm{dt}} = \frac{n}{12} K_{\mathrm{inel}} T_z T_x - \frac{\Gamma_{th}}{3} (T_z-T_x) + c_l \, , \label{eqn:ode_elastic_inelastic2}\\
&\frac{\mathrm{d}T_x}{\mathrm{dt}} = \frac{n}{12} K_{\mathrm{inel}} (2T_z-T_x) T_x + \frac{2 \Gamma_{th}}{3} (T_z-T_x) + c_l \, \label{eqn:ode_elastic_inelastic3}. 
\end{align}
Here, $n$ is the average density of the sample and $K_\text{inel}= \beta_\text{inel}/T$ is the temperature-independent coefficient of inelastic collisions. $T_x$ and $T_z$ are the effective temperatures of the molecules in the horizontal and vertical direction, respectively. We checked that the temperature along the direction of the imaging beam $T_y$ is equal to $T_z$ and assume this for all measurements. $\Gamma_{th}$ is the thermalization rate given by $\Gamma_{th} = n \sigma \bar{v}/\alpha$ where $\sigma$ is the cross-section of the elastic collisions which is assumed to be constant for each measurement, $\bar{v}=\sqrt{16 k_B (2T_z+T_x)/(3 \pi m)}$ is the average velocity of molecules and $\alpha = 4.1$ is the number of collisions needed for thermalization in $p$-wave collisions \cite{DeMarco1999,Ni2010}. The linear heating term $c_l$ has been introduced to phenomenologically account for isotropic heating that we observe in these measurements. 
We numerically fit Eqs.~\eqref{eqn:ode_elastic_inelastic1}--\eqref{eqn:ode_elastic_inelastic3} in the basis of the average temperature $T_{\mathrm{avg}}=(2T_z+T_x)/3$ and the difference in temperature $\Delta T=T_z-T_x$
to reduce the common mode fluctuation in the temperatures. 

\begin{figure}
\centering
\includegraphics{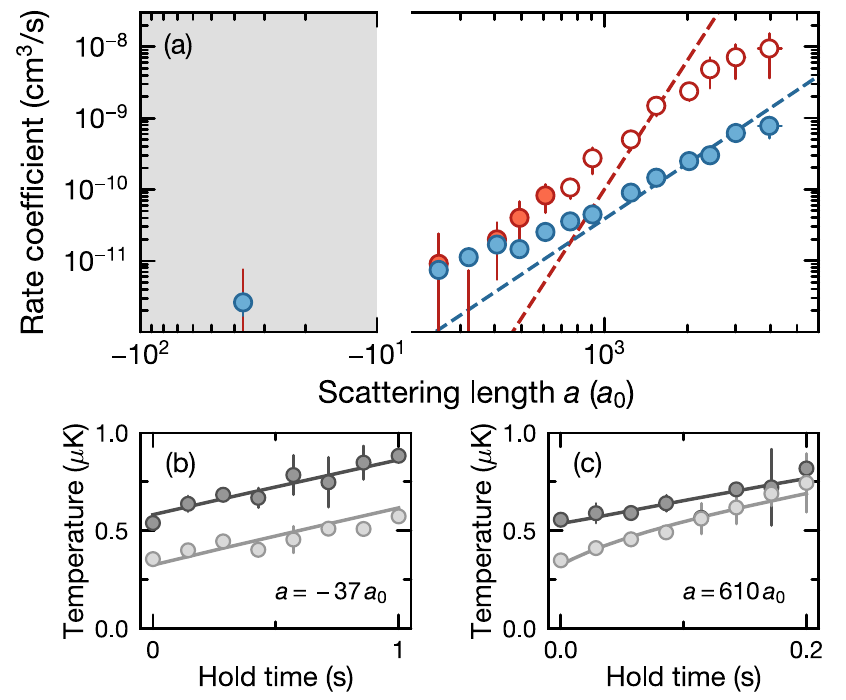}
\caption{Elastic $p$-wave collisions of fermionic Feshbach molecules. (a) Elastic collision coefficient $\beta_{\mathrm{el}}$ (red) and inelastic collision coefficient $\beta_{\mathrm{inel}}$ (blue) as a function of the scattering length $a$ for a temperature $T=\SI{500}{nK}$. Data points for $\beta_{\mathrm{el}}$ are marked (open symbols) when the number of K atoms increases during the measurement. The red dashed line shows an $a^6$ scaling and serves as a guide to the eye. The blue dashed line shows the $a^{2.58}$ scaling which is deduced from a fit to the loss coefficients in the first set of measurement. The vertical error bars for the inelastic and elastic collision rates are given by the error of the fit. The horizontal error bars result from a 15-mG uncertainty in the magnetic field. A systematic error of 30\% in the collision rate coefficients is expected from a 10\% error in the trapping frequency. The rate coefficient for elastic collisions for $a=-37\,a_0$ and $322\,a_0$ lie outside of the plotting range. (b),(c) Effective temperatures $T_z$ (dark gray) and $T_x$ (light gray) as a function of the hold time for magnetic fields of \SI{72.25}{G} ($-37\,a_0$) and \SI{77.25}{G} ($610\,a_0$), respectively.}
\label{fig:inelastic_elastic}
\end{figure}

The resulting coefficients of inelastic and elastic collisions, $\beta_{\mathrm{inel}}$ and $\beta_{\mathrm{el}}=\sigma \bar{v}$, are summarized in Fig.~\ref{fig:inelastic_elastic} as a function of the scattering length for a temperature of \SI{500}{nK}. Deep in the non-halo regime, collisions of Feshbach molecules predominantly result in loss such that the number of elastic collisions during the lifetime of the molecular sample is negligible. As a result, cross-dimensional thermalization is absent (see Fig.~\ref{fig:inelastic_elastic}b). For higher scattering lengths the elastic collision rate increases notably faster than the inelastic collision rate. Due to cross-dimensional thermalization, the temperatures in the two directions approach each other during the measurement (see Fig.~\ref{fig:inelastic_elastic}c).

Unfortunately, our measurement does not allow us to extract the scaling of $\beta_\text{el}$ with the scattering length $a$ in the halo regime due to the following two reasons. First, while we intentionally reduced the initial molecule density to $n_0=0.4 \times 10^{11}\,\mathrm{cm^{-3}}$, the elastic collision rate becomes comparable to the trap frequencies for a scattering length of $a>2000\,a_0$. Therefore the system enters the hydrodynamic regime and the measured value of $\beta_\text{el}$ saturates near the so-called hydrodynamic limit given by $\bar{\omega} \alpha/(2 \pi n_0)=1.71 \times 10^{-9}\,\mathrm{cm^3/s}$ with a geometric mean trapping frequency $\bar{\omega}=2 \pi \times 167\,\mathrm{Hz}$ \cite{Ma2003}. We note that in the presence of strong loss, the measured elastic collision rate can exceed the hydrodynamic limit calculated for a constant density. Second, our model which only considers dimer-dimer elastic collisions, is only valid in deducing $\beta_\text{el}$ for scattering lengths up to $610\,a_0$. When $a > 610\,a_0$, where the binding energy of the molecules is smaller than the trap depth, we observe an accumulation of spectator K atoms which can contribute to the cross-thermalization. Despite these limitations, this measurement shows that the rate coefficient of elastic collisions exhibits a stronger scaling with the scattering length than that of inelastic collisions.

\begin{figure}
\centering
\includegraphics{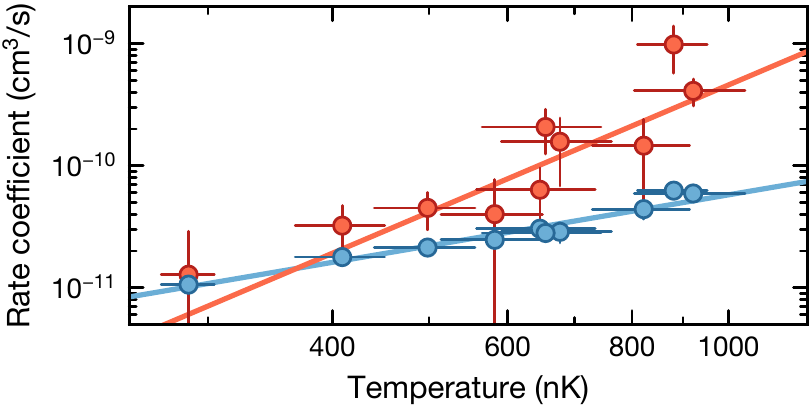}
\caption{Elastic $p$-wave collisions in fermionic Feshbach molecules for different temperatures. Elastic collision coefficient $\beta_{\mathrm{el}}$ (red) and inelastic collision coefficient $\beta_{\mathrm{inel}}$ (blue) as a function of the temperature averaged during the hold time for a scattering length $a=610\,a_0$. The vertical error bars are given by the error of the fit. The horizontal error bars represent the statistical error of $T_\text{avg}$ during the hold time. A systematic error of 30\% in the collision rate is expected from a 10\% error in the trapping frequency. The blue and red solid line are the fits of the polynomial function to the rate coefficients.}
\label{fig:inelastic_elastic_T}
\end{figure}

In addition, we measure the scaling of the elastic collision rate with the temperature. To this end, we perform the non-adiabatic compression to different trap depths while keeping the ratio between the trapping frequencies fixed. To ensure that cross-dimensional thermalization results from elastic collisions between the Feshbach molecules, we perform the measurement at a scattering length of $a=610\,a_0$ where we checked that the number of K atoms does not systematically change for the different temperatures. The results are summarized in Fig.~\ref{fig:inelastic_elastic_T}. One can see that the elastic rate coefficient scales stronger with temperature compared to the inelastic one. For temperatures around $T=\SI{300}{nK}$, the rate of elastic and inelastic collisions are comparable while for $T=\SI{1000}{nK}$, the elastic collision rate is larger by almost one order of magnitude. We fit the temperature scaling with the polynomial $c T^l$ and obtain $l=1.39(12)$ and $l=3.46(63)$ for $\beta_{\mathrm{inel}}$ and $\beta_{\mathrm{el}}$, respectively. The measured temperature scaling of the elastic collision rate agrees reasonably with the $T^{5/2}$ scaling expected for $p$-wave collisions.

In conclusion, we have characterized inelastic and elastic collisions of fermionic $\mathrm{NaK}$ Feshbach molecules and found reasonable agreement with the threshold behavior of $p$-wave collisions between Feshbach molecules. Our investigation provides a benchmark for solving the four-body problem that predicts the collisional behavior in other heteronuclear molecules. The surprisingly strong suppression of dimer-dimer loss in the non-halo regime suggests that after molecule formation, dimer loss can be strongly reduced by quenching the magnetic field into the regime of large binding energies. This has, for example, been essential in our recent work on the generation of degenerate Fermi gases of $\mathrm{NaK}$ molecules \cite{Duda2021,Schindewolf2022}.

With the discovery of long-lived interacting Feshbach molecules, it seems unfortunate that the pursuit for utilizing these molecules for quantum many-body physics has predominantly ceased. Further improvement of the ratio of elastic-to-inelastic collisions is expected when confining the molecules in two-dimensional traps \cite{Bazak2018, Zhang2021}, which should allow for evaporative cooling of Feshbach molecules and possibly the observation of $p$-wave paired phases. The authors of Ref.~\cite{Bazak2018} have identified \text{$^{40}$K--$^{39}$K} as a promising system in this regard. Furthermore, it would be interesting to prepare fermionic Feshbach molecules formed by two magnetic atoms such as $^{170}\text{Er}^{161}\text{Dy}$ \cite{Durastante2020}, which are expected to be long-lived and exhibit large magnetic dipole moments \cite{Frisch2015}.

\begin{acknowledgments}
We thank D. Petrov for discussions on the two-body loss mechanism and stimulating discussions. We gratefully acknowledge support from the Max Planck Society, the European Union (PASQuanS Grant No.\ 817482) and the Deutsche Forschungsgemeinschaft under Germany's Excellence Strategy – EXC-2111 – 390814868 and under Grant No.\ FOR 2247. A.S.\ acknowledges funding from the Max Planck Harvard Research Center for Quantum Optics.
\end{acknowledgments}

\bibliography{bibliography}

\begin{thebibliography}{33}%
\makeatletter
\providecommand \@ifxundefined [1]{%
 \@ifx{#1\undefined}
}%
\providecommand \@ifnum [1]{%
 \ifnum #1\expandafter \@firstoftwo
 \else \expandafter \@secondoftwo
 \fi
}%
\providecommand \@ifx [1]{%
 \ifx #1\expandafter \@firstoftwo
 \else \expandafter \@secondoftwo
 \fi
}%
\providecommand \natexlab [1]{#1}%
\providecommand \enquote  [1]{``#1''}%
\providecommand \bibnamefont  [1]{#1}%
\providecommand \bibfnamefont [1]{#1}%
\providecommand \citenamefont [1]{#1}%
\providecommand \href@noop [0]{\@secondoftwo}%
\providecommand \href [0]{\begingroup \@sanitize@url \@href}%
\providecommand \@href[1]{\@@startlink{#1}\@@href}%
\providecommand \@@href[1]{\endgroup#1\@@endlink}%
\providecommand \@sanitize@url [0]{\catcode `\\12\catcode `\$12\catcode
  `\&12\catcode `\#12\catcode `\^12\catcode `\_12\catcode `\%12\relax}%
\providecommand \@@startlink[1]{}%
\providecommand \@@endlink[0]{}%
\providecommand \url  [0]{\begingroup\@sanitize@url \@url }%
\providecommand \@url [1]{\endgroup\@href {#1}{\urlprefix }}%
\providecommand \urlprefix  [0]{URL }%
\providecommand \Eprint [0]{\href }%
\providecommand \doibase [0]{http://dx.doi.org/}%
\providecommand \selectlanguage [0]{\@gobble}%
\providecommand \bibinfo  [0]{\@secondoftwo}%
\providecommand \bibfield  [0]{\@secondoftwo}%
\providecommand \translation [1]{[#1]}%
\providecommand \BibitemOpen [0]{}%
\providecommand \bibitemStop [0]{}%
\providecommand \bibitemNoStop [0]{.\EOS\space}%
\providecommand \EOS [0]{\spacefactor3000\relax}%
\providecommand \BibitemShut  [1]{\csname bibitem#1\endcsname}%
\let\auto@bib@innerbib\@empty
\bibitem [{\citenamefont {Carr}\ \emph {et~al.}(2009)\citenamefont {Carr},
  \citenamefont {DeMille}, \citenamefont {Krems},\ and\ \citenamefont
  {Ye}}]{Carr2009}%
  \BibitemOpen
  \bibfield  {author} {\bibinfo {author} {\bibfnamefont {L.~D.}\ \bibnamefont
  {Carr}}, \bibinfo {author} {\bibfnamefont {D.}~\bibnamefont {DeMille}},
  \bibinfo {author} {\bibfnamefont {R.~V.}\ \bibnamefont {Krems}}, \ and\
  \bibinfo {author} {\bibfnamefont {J.}~\bibnamefont {Ye}},\ }\href {\doibase
  10.1088/1367-2630/11/5/055049} {\bibfield  {journal} {\bibinfo  {journal}
  {New J. Phys.}\ }\textbf {\bibinfo {volume} {11}},\ \bibinfo {pages} {055049}
  (\bibinfo {year} {2009})}\BibitemShut {NoStop}%
\bibitem [{\citenamefont {Qu{\'e}m{\'e}ner}\ and\ \citenamefont
  {Julienne}(2012)}]{Quemener2012}%
  \BibitemOpen
  \bibfield  {author} {\bibinfo {author} {\bibfnamefont {G.}~\bibnamefont
  {Qu{\'e}m{\'e}ner}}\ and\ \bibinfo {author} {\bibfnamefont {P.~S.}\
  \bibnamefont {Julienne}},\ }\href {\doibase 10.1021/cr300092g} {\bibfield
  {journal} {\bibinfo  {journal} {Chem. Rev.}\ }\textbf {\bibinfo {volume}
  {112}},\ \bibinfo {pages} {4949} (\bibinfo {year} {2012})}\BibitemShut
  {NoStop}%
\bibitem [{\citenamefont {Bohn}\ \emph {et~al.}(2017)\citenamefont {Bohn},
  \citenamefont {Rey},\ and\ \citenamefont {Ye}}]{Bohn2017}%
  \BibitemOpen
  \bibfield  {author} {\bibinfo {author} {\bibfnamefont {J.~L.}\ \bibnamefont
  {Bohn}}, \bibinfo {author} {\bibfnamefont {A.~M.}\ \bibnamefont {Rey}}, \
  and\ \bibinfo {author} {\bibfnamefont {J.}~\bibnamefont {Ye}},\ }\href
  {\doibase 10.1126/science.aam6299} {\bibfield  {journal} {\bibinfo  {journal}
  {Science}\ }\textbf {\bibinfo {volume} {357}},\ \bibinfo {pages} {1002}
  (\bibinfo {year} {2017})}\BibitemShut {NoStop}%
\bibitem [{\citenamefont {Chin}\ \emph {et~al.}(2010)\citenamefont {Chin},
  \citenamefont {Grimm}, \citenamefont {Julienne},\ and\ \citenamefont
  {Tiesinga}}]{Chin2010}%
  \BibitemOpen
  \bibfield  {author} {\bibinfo {author} {\bibfnamefont {C.}~\bibnamefont
  {Chin}}, \bibinfo {author} {\bibfnamefont {R.}~\bibnamefont {Grimm}},
  \bibinfo {author} {\bibfnamefont {P.}~\bibnamefont {Julienne}}, \ and\
  \bibinfo {author} {\bibfnamefont {E.}~\bibnamefont {Tiesinga}},\ }\href
  {\doibase 10.1103/RevModPhys.82.1225} {\bibfield  {journal} {\bibinfo
  {journal} {Rev. Mod. Phys.}\ }\textbf {\bibinfo {volume} {82}},\ \bibinfo
  {pages} {1225} (\bibinfo {year} {2010})}\BibitemShut {NoStop}%
\bibitem [{\citenamefont {Ferlaino}\ \emph {et~al.}(2008)\citenamefont
  {Ferlaino}, \citenamefont {Knoop}, \citenamefont {Mark}, \citenamefont
  {Berninger}, \citenamefont {Sch{\"o}bel}, \citenamefont {N{\"a}gerl},\ and\
  \citenamefont {Grimm}}]{Ferlaino2008}%
  \BibitemOpen
  \bibfield  {author} {\bibinfo {author} {\bibfnamefont {F.}~\bibnamefont
  {Ferlaino}}, \bibinfo {author} {\bibfnamefont {S.}~\bibnamefont {Knoop}},
  \bibinfo {author} {\bibfnamefont {M.}~\bibnamefont {Mark}}, \bibinfo {author}
  {\bibfnamefont {M.}~\bibnamefont {Berninger}}, \bibinfo {author}
  {\bibfnamefont {H.}~\bibnamefont {Sch{\"o}bel}}, \bibinfo {author}
  {\bibfnamefont {H.-C.}\ \bibnamefont {N{\"a}gerl}}, \ and\ \bibinfo {author}
  {\bibfnamefont {R.}~\bibnamefont {Grimm}},\ }\href {\doibase
  10.1103/PhysRevLett.101.023201} {\bibfield  {journal} {\bibinfo  {journal}
  {Phys. Rev. Lett.}\ }\textbf {\bibinfo {volume} {101}},\ \bibinfo {pages}
  {023201} (\bibinfo {year} {2008})}\BibitemShut {NoStop}%
\bibitem [{\citenamefont {Ospelkaus}\ \emph {et~al.}(2010)\citenamefont
  {Ospelkaus}, \citenamefont {Ni}, \citenamefont {Wang}, \citenamefont
  {de~Miranda}, \citenamefont {Neyenhuis}, \citenamefont {Qu{\'e}m{\'e}ner},
  \citenamefont {Julienne}, \citenamefont {Bohn}, \citenamefont {Jin},\ and\
  \citenamefont {Ye}}]{Ospelkaus2010}%
  \BibitemOpen
  \bibfield  {author} {\bibinfo {author} {\bibfnamefont {S.}~\bibnamefont
  {Ospelkaus}}, \bibinfo {author} {\bibfnamefont {K.-K.}\ \bibnamefont {Ni}},
  \bibinfo {author} {\bibfnamefont {D.}~\bibnamefont {Wang}}, \bibinfo {author}
  {\bibfnamefont {M.~H. G.~d.}\ \bibnamefont {de~Miranda}}, \bibinfo {author}
  {\bibfnamefont {B.}~\bibnamefont {Neyenhuis}}, \bibinfo {author}
  {\bibfnamefont {G.}~\bibnamefont {Qu{\'e}m{\'e}ner}}, \bibinfo {author}
  {\bibfnamefont {P.~S.}\ \bibnamefont {Julienne}}, \bibinfo {author}
  {\bibfnamefont {J.~L.}\ \bibnamefont {Bohn}}, \bibinfo {author}
  {\bibfnamefont {D.~S.}\ \bibnamefont {Jin}}, \ and\ \bibinfo {author}
  {\bibfnamefont {J.}~\bibnamefont {Ye}},\ }\href {\doibase
  10.1126/science.1184121} {\bibfield  {journal} {\bibinfo  {journal}
  {Science}\ }\textbf {\bibinfo {volume} {327}},\ \bibinfo {pages} {853}
  (\bibinfo {year} {2010})}\BibitemShut {NoStop}%
\bibitem [{\citenamefont {Bause}\ \emph
  {et~al.}(2021{\natexlab{a}})\citenamefont {Bause}, \citenamefont {Kamijo},
  \citenamefont {Chen}, \citenamefont {Duda}, \citenamefont {Schindewolf},
  \citenamefont {Bloch},\ and\ \citenamefont {Luo}}]{Bause2021b}%
  \BibitemOpen
  \bibfield  {author} {\bibinfo {author} {\bibfnamefont {R.}~\bibnamefont
  {Bause}}, \bibinfo {author} {\bibfnamefont {A.}~\bibnamefont {Kamijo}},
  \bibinfo {author} {\bibfnamefont {X.-Y.}\ \bibnamefont {Chen}}, \bibinfo
  {author} {\bibfnamefont {M.}~\bibnamefont {Duda}}, \bibinfo {author}
  {\bibfnamefont {A.}~\bibnamefont {Schindewolf}}, \bibinfo {author}
  {\bibfnamefont {I.}~\bibnamefont {Bloch}}, \ and\ \bibinfo {author}
  {\bibfnamefont {X.-Y.}\ \bibnamefont {Luo}},\ }\href {\doibase
  10.1103/PhysRevA.104.043321} {\bibfield  {journal} {\bibinfo  {journal}
  {Phys. Rev. A}\ }\textbf {\bibinfo {volume} {104}},\ \bibinfo {pages}
  {043321} (\bibinfo {year} {2021}{\natexlab{a}})}\BibitemShut {NoStop}%
\bibitem [{\citenamefont {Marcelis}\ \emph {et~al.}(2008)\citenamefont
  {Marcelis}, \citenamefont {Kokkelmans}, \citenamefont {Shlyapnikov},\ and\
  \citenamefont {Petrov}}]{Marcelis2008}%
  \BibitemOpen
  \bibfield  {author} {\bibinfo {author} {\bibfnamefont {B.}~\bibnamefont
  {Marcelis}}, \bibinfo {author} {\bibfnamefont {S.~J. J. M.~F.}\ \bibnamefont
  {Kokkelmans}}, \bibinfo {author} {\bibfnamefont {G.~V.}\ \bibnamefont
  {Shlyapnikov}}, \ and\ \bibinfo {author} {\bibfnamefont {D.~S.}\ \bibnamefont
  {Petrov}},\ }\href {\doibase 10.1103/PhysRevA.77.032707} {\bibfield
  {journal} {\bibinfo  {journal} {Phys. Rev. A}\ }\textbf {\bibinfo {volume}
  {77}},\ \bibinfo {pages} {032707} (\bibinfo {year} {2008})}\BibitemShut
  {NoStop}%
\bibitem [{\citenamefont {Maeda}\ \emph {et~al.}(2009)\citenamefont {Maeda},
  \citenamefont {Baym},\ and\ \citenamefont {Hatsuda}}]{Kenji2009}%
  \BibitemOpen
  \bibfield  {author} {\bibinfo {author} {\bibfnamefont {K.}~\bibnamefont
  {Maeda}}, \bibinfo {author} {\bibfnamefont {G.}~\bibnamefont {Baym}}, \ and\
  \bibinfo {author} {\bibfnamefont {T.}~\bibnamefont {Hatsuda}},\ }\href
  {\doibase 10.1103/PhysRevLett.103.085301} {\bibfield  {journal} {\bibinfo
  {journal} {Phys. Rev. Lett.}\ }\textbf {\bibinfo {volume} {103}},\ \bibinfo
  {pages} {085301} (\bibinfo {year} {2009})}\BibitemShut {NoStop}%
\bibitem [{\citenamefont {Bazak}\ and\ \citenamefont
  {Petrov}(2018)}]{Bazak2018}%
  \BibitemOpen
  \bibfield  {author} {\bibinfo {author} {\bibfnamefont {B.}~\bibnamefont
  {Bazak}}\ and\ \bibinfo {author} {\bibfnamefont {D.~S.}\ \bibnamefont
  {Petrov}},\ }\href {\doibase 10.1103/PhysRevLett.121.263001} {\bibfield
  {journal} {\bibinfo  {journal} {Phys. Rev. Lett.}\ }\textbf {\bibinfo
  {volume} {121}},\ \bibinfo {pages} {263001} (\bibinfo {year}
  {2018})}\BibitemShut {NoStop}%
\bibitem [{\citenamefont {Greiner}\ \emph {et~al.}(2003)\citenamefont
  {Greiner}, \citenamefont {Regal},\ and\ \citenamefont {Jin}}]{Greiner2003}%
  \BibitemOpen
  \bibfield  {author} {\bibinfo {author} {\bibfnamefont {M.}~\bibnamefont
  {Greiner}}, \bibinfo {author} {\bibfnamefont {C.~A.}\ \bibnamefont {Regal}},
  \ and\ \bibinfo {author} {\bibfnamefont {D.~S.}\ \bibnamefont {Jin}},\ }\href
  {\doibase 10.1038/nature02199} {\bibfield  {journal} {\bibinfo  {journal}
  {Nature}\ }\textbf {\bibinfo {volume} {426}},\ \bibinfo {pages} {537}
  (\bibinfo {year} {2003})}\BibitemShut {NoStop}%
\bibitem [{\citenamefont {Regal}\ \emph {et~al.}(2004)\citenamefont {Regal},
  \citenamefont {Greiner},\ and\ \citenamefont {Jin}}]{Regal2004}%
  \BibitemOpen
  \bibfield  {author} {\bibinfo {author} {\bibfnamefont {C.~A.}\ \bibnamefont
  {Regal}}, \bibinfo {author} {\bibfnamefont {M.}~\bibnamefont {Greiner}}, \
  and\ \bibinfo {author} {\bibfnamefont {D.~S.}\ \bibnamefont {Jin}},\ }\href
  {\doibase 10.1103/PhysRevLett.92.040403} {\bibfield  {journal} {\bibinfo
  {journal} {Phys. Rev. Lett.}\ }\textbf {\bibinfo {volume} {92}},\ \bibinfo
  {pages} {040403} (\bibinfo {year} {2004})}\BibitemShut {NoStop}%
\bibitem [{\citenamefont {Zirbel}\ \emph {et~al.}(2008)\citenamefont {Zirbel},
  \citenamefont {Ni}, \citenamefont {Ospelkaus}, \citenamefont
  {D{\textquoteright}Incao}, \citenamefont {Wieman}, \citenamefont {Ye},\ and\
  \citenamefont {Jin}}]{Zirbel2008}%
  \BibitemOpen
  \bibfield  {author} {\bibinfo {author} {\bibfnamefont {J.~J.}\ \bibnamefont
  {Zirbel}}, \bibinfo {author} {\bibfnamefont {K.-K.}\ \bibnamefont {Ni}},
  \bibinfo {author} {\bibfnamefont {S.}~\bibnamefont {Ospelkaus}}, \bibinfo
  {author} {\bibfnamefont {J.~P.}\ \bibnamefont {D{\textquoteright}Incao}},
  \bibinfo {author} {\bibfnamefont {C.~E.}\ \bibnamefont {Wieman}}, \bibinfo
  {author} {\bibfnamefont {J.}~\bibnamefont {Ye}}, \ and\ \bibinfo {author}
  {\bibfnamefont {D.~S.}\ \bibnamefont {Jin}},\ }\href {\doibase
  10.1103/PhysRevLett.100.143201} {\bibfield  {journal} {\bibinfo  {journal}
  {Phys. Rev. Lett.}\ }\textbf {\bibinfo {volume} {100}},\ \bibinfo {pages}
  {143201} (\bibinfo {year} {2008})}\BibitemShut {NoStop}%
\bibitem [{\citenamefont {Wu}\ \emph {et~al.}(2012)\citenamefont {Wu},
  \citenamefont {Park}, \citenamefont {Ahmadi}, \citenamefont {Will},\ and\
  \citenamefont {Zwierlein}}]{Wu2012}%
  \BibitemOpen
  \bibfield  {author} {\bibinfo {author} {\bibfnamefont {C.-H.}\ \bibnamefont
  {Wu}}, \bibinfo {author} {\bibfnamefont {J.~W.}\ \bibnamefont {Park}},
  \bibinfo {author} {\bibfnamefont {P.}~\bibnamefont {Ahmadi}}, \bibinfo
  {author} {\bibfnamefont {S.}~\bibnamefont {Will}}, \ and\ \bibinfo {author}
  {\bibfnamefont {M.~W.}\ \bibnamefont {Zwierlein}},\ }\href {\doibase
  10.1103/PhysRevLett.109.085301} {\bibfield  {journal} {\bibinfo  {journal}
  {Phys. Rev. Lett.}\ }\textbf {\bibinfo {volume} {109}},\ \bibinfo {pages}
  {085301} (\bibinfo {year} {2012})}\BibitemShut {NoStop}%
\bibitem [{\citenamefont {Heo}\ \emph {et~al.}(2012)\citenamefont {Heo},
  \citenamefont {Wang}, \citenamefont {Christensen}, \citenamefont {Rvachov},
  \citenamefont {Cotta}, \citenamefont {Choi}, \citenamefont {Lee},\ and\
  \citenamefont {Ketterle}}]{Heo2012}%
  \BibitemOpen
  \bibfield  {author} {\bibinfo {author} {\bibfnamefont {M.-S.}\ \bibnamefont
  {Heo}}, \bibinfo {author} {\bibfnamefont {T.~T.}\ \bibnamefont {Wang}},
  \bibinfo {author} {\bibfnamefont {C.~A.}\ \bibnamefont {Christensen}},
  \bibinfo {author} {\bibfnamefont {T.~M.}\ \bibnamefont {Rvachov}}, \bibinfo
  {author} {\bibfnamefont {D.~A.}\ \bibnamefont {Cotta}}, \bibinfo {author}
  {\bibfnamefont {J.-H.}\ \bibnamefont {Choi}}, \bibinfo {author}
  {\bibfnamefont {Y.-R.}\ \bibnamefont {Lee}}, \ and\ \bibinfo {author}
  {\bibfnamefont {W.}~\bibnamefont {Ketterle}},\ }\href {\doibase
  10.1103/PhysRevA.86.021602} {\bibfield  {journal} {\bibinfo  {journal} {Phys.
  Rev. A}\ }\textbf {\bibinfo {volume} {86}},\ \bibinfo {pages} {021602}
  (\bibinfo {year} {2012})}\BibitemShut {NoStop}%
\bibitem [{\citenamefont {Petrov}(2020)}]{privateDima2020}%
  \BibitemOpen
  \bibfield  {author} {\bibinfo {author} {\bibfnamefont {D.~S.}\ \bibnamefont
  {Petrov}},\ }\href@noop {} {\bibfield  {journal} {\bibinfo  {journal}
  {private communication}\ } (\bibinfo {year} {2020})}\BibitemShut {NoStop}%
\bibitem [{\citenamefont {Chen}\ \emph {et~al.}(2021)\citenamefont {Chen},
  \citenamefont {Duda}, \citenamefont {Schindewolf}, \citenamefont {Bause},
  \citenamefont {Bloch},\ and\ \citenamefont {Luo}}]{Chen2021}%
  \BibitemOpen
  \bibfield  {author} {\bibinfo {author} {\bibfnamefont {X.-Y.}\ \bibnamefont
  {Chen}}, \bibinfo {author} {\bibfnamefont {M.}~\bibnamefont {Duda}}, \bibinfo
  {author} {\bibfnamefont {A.}~\bibnamefont {Schindewolf}}, \bibinfo {author}
  {\bibfnamefont {R.}~\bibnamefont {Bause}}, \bibinfo {author} {\bibfnamefont
  {I.}~\bibnamefont {Bloch}}, \ and\ \bibinfo {author} {\bibfnamefont {X.-Y.}\
  \bibnamefont {Luo}},\ }\href {http://arxiv.org/abs/2110.01290} {\bibfield
  {journal} {\bibinfo  {journal} {arXiv:2110.01290}\ } (\bibinfo {year}
  {2021})}\BibitemShut {NoStop}%
\bibitem [{\citenamefont {Viel}\ and\ \citenamefont {Simoni}(2016)}]{Viel2016}%
  \BibitemOpen
  \bibfield  {author} {\bibinfo {author} {\bibfnamefont {A.}~\bibnamefont
  {Viel}}\ and\ \bibinfo {author} {\bibfnamefont {A.}~\bibnamefont {Simoni}},\
  }\href {\doibase 10.1103/PhysRevA.93.042701} {\bibfield  {journal} {\bibinfo
  {journal} {Phys. Rev. A}\ }\textbf {\bibinfo {volume} {93}},\ \bibinfo
  {pages} {042701} (\bibinfo {year} {2016})}\BibitemShut {NoStop}%
\bibitem [{\citenamefont {Bause}\ \emph
  {et~al.}(2021{\natexlab{b}})\citenamefont {Bause}, \citenamefont
  {Schindewolf}, \citenamefont {Tao}, \citenamefont {Duda}, \citenamefont
  {Chen}, \citenamefont {Qu{\'e}m{\'e}ner}, \citenamefont {Karman},
  \citenamefont {Christianen}, \citenamefont {Bloch},\ and\ \citenamefont
  {Luo}}]{Bause2021}%
  \BibitemOpen
  \bibfield  {author} {\bibinfo {author} {\bibfnamefont {R.}~\bibnamefont
  {Bause}}, \bibinfo {author} {\bibfnamefont {A.}~\bibnamefont {Schindewolf}},
  \bibinfo {author} {\bibfnamefont {R.}~\bibnamefont {Tao}}, \bibinfo {author}
  {\bibfnamefont {M.}~\bibnamefont {Duda}}, \bibinfo {author} {\bibfnamefont
  {X.-Y.}\ \bibnamefont {Chen}}, \bibinfo {author} {\bibfnamefont
  {G.}~\bibnamefont {Qu{\'e}m{\'e}ner}}, \bibinfo {author} {\bibfnamefont
  {T.}~\bibnamefont {Karman}}, \bibinfo {author} {\bibfnamefont
  {A.}~\bibnamefont {Christianen}}, \bibinfo {author} {\bibfnamefont
  {I.}~\bibnamefont {Bloch}}, \ and\ \bibinfo {author} {\bibfnamefont {X.-Y.}\
  \bibnamefont {Luo}},\ }\href {\doibase 10.1103/PhysRevResearch.3.033013}
  {\bibfield  {journal} {\bibinfo  {journal} {Phys. Rev. Research}\ }\textbf
  {\bibinfo {volume} {3}},\ \bibinfo {pages} {033013} (\bibinfo {year}
  {2021}{\natexlab{b}})}\BibitemShut {NoStop}%
\bibitem [{\citenamefont {Yan}\ \emph {et~al.}(2020)\citenamefont {Yan},
  \citenamefont {Park}, \citenamefont {Ni}, \citenamefont {Loh}, \citenamefont
  {Will}, \citenamefont {Karman},\ and\ \citenamefont {Zwierlein}}]{Yan2020}%
  \BibitemOpen
  \bibfield  {author} {\bibinfo {author} {\bibfnamefont {Z.~Z.}\ \bibnamefont
  {Yan}}, \bibinfo {author} {\bibfnamefont {J.~W.}\ \bibnamefont {Park}},
  \bibinfo {author} {\bibfnamefont {Y.}~\bibnamefont {Ni}}, \bibinfo {author}
  {\bibfnamefont {H.}~\bibnamefont {Loh}}, \bibinfo {author} {\bibfnamefont
  {S.}~\bibnamefont {Will}}, \bibinfo {author} {\bibfnamefont {T.}~\bibnamefont
  {Karman}}, \ and\ \bibinfo {author} {\bibfnamefont {M.}~\bibnamefont
  {Zwierlein}},\ }\href {\doibase 10.1103/PhysRevLett.125.063401} {\bibfield
  {journal} {\bibinfo  {journal} {Phys. Rev. Lett.}\ }\textbf {\bibinfo
  {volume} {125}},\ \bibinfo {pages} {063401} (\bibinfo {year}
  {2020})}\BibitemShut {NoStop}%
\bibitem [{\citenamefont {Hudson}\ \emph {et~al.}(2008)\citenamefont {Hudson},
  \citenamefont {Gilfoy}, \citenamefont {Kotochigova}, \citenamefont {Sage},\
  and\ \citenamefont {DeMille}}]{Hudson2008}%
  \BibitemOpen
  \bibfield  {author} {\bibinfo {author} {\bibfnamefont {E.~R.}\ \bibnamefont
  {Hudson}}, \bibinfo {author} {\bibfnamefont {N.~B.}\ \bibnamefont {Gilfoy}},
  \bibinfo {author} {\bibfnamefont {S.}~\bibnamefont {Kotochigova}}, \bibinfo
  {author} {\bibfnamefont {J.~M.}\ \bibnamefont {Sage}}, \ and\ \bibinfo
  {author} {\bibfnamefont {D.}~\bibnamefont {DeMille}},\ }\href {\doibase
  10.1103/PhysRevLett.100.203201} {\bibfield  {journal} {\bibinfo  {journal}
  {Phys. Rev. Lett.}\ }\textbf {\bibinfo {volume} {100}},\ \bibinfo {pages}
  {203201} (\bibinfo {year} {2008})}\BibitemShut {NoStop}%
\bibitem [{\citenamefont {Idziaszek}\ and\ \citenamefont
  {Julienne}(2010)}]{Idziaszek2010}%
  \BibitemOpen
  \bibfield  {author} {\bibinfo {author} {\bibfnamefont {Z.}~\bibnamefont
  {Idziaszek}}\ and\ \bibinfo {author} {\bibfnamefont {P.~S.}\ \bibnamefont
  {Julienne}},\ }\href {\doibase 10.1103/PhysRevLett.104.113202} {\bibfield
  {journal} {\bibinfo  {journal} {Phys. Rev. Lett.}\ }\textbf {\bibinfo
  {volume} {104}},\ \bibinfo {pages} {113202} (\bibinfo {year}
  {2010})}\BibitemShut {NoStop}%
\bibitem [{\citenamefont {Derevianko}\ \emph {et~al.}(2001)\citenamefont
  {Derevianko}, \citenamefont {Babb},\ and\ \citenamefont
  {Dalgarno}}]{Derevianko2001}%
  \BibitemOpen
  \bibfield  {author} {\bibinfo {author} {\bibfnamefont {A.}~\bibnamefont
  {Derevianko}}, \bibinfo {author} {\bibfnamefont {J.~F.}\ \bibnamefont
  {Babb}}, \ and\ \bibinfo {author} {\bibfnamefont {A.}~\bibnamefont
  {Dalgarno}},\ }\href {\doibase 10.1103/PhysRevA.63.052704} {\bibfield
  {journal} {\bibinfo  {journal} {Phys. Rev. A}\ }\textbf {\bibinfo {volume}
  {63}},\ \bibinfo {pages} {052704} (\bibinfo {year} {2001})}\BibitemShut
  {NoStop}%
\bibitem [{\citenamefont {Mitroy}\ and\ \citenamefont
  {Bromley}(2003)}]{Mitroy2003}%
  \BibitemOpen
  \bibfield  {author} {\bibinfo {author} {\bibfnamefont {J.}~\bibnamefont
  {Mitroy}}\ and\ \bibinfo {author} {\bibfnamefont {M.~W.~J.}\ \bibnamefont
  {Bromley}},\ }\href {\doibase 10.1103/PhysRevA.68.052714} {\bibfield
  {journal} {\bibinfo  {journal} {Phys. Rev. A}\ }\textbf {\bibinfo {volume}
  {68}},\ \bibinfo {pages} {052714} (\bibinfo {year} {2003})}\BibitemShut
  {NoStop}%
\bibitem [{\citenamefont {Julienne}\ \emph {et~al.}(2011)\citenamefont
  {Julienne}, \citenamefont {Hanna},\ and\ \citenamefont
  {Idziaszek}}]{Julienne2011}%
  \BibitemOpen
  \bibfield  {author} {\bibinfo {author} {\bibfnamefont {P.~S.}\ \bibnamefont
  {Julienne}}, \bibinfo {author} {\bibfnamefont {T.~M.}\ \bibnamefont {Hanna}},
  \ and\ \bibinfo {author} {\bibfnamefont {Z.}~\bibnamefont {Idziaszek}},\
  }\href {\doibase 10.1039/C1CP21270B} {\bibfield  {journal} {\bibinfo
  {journal} {Phys. Chem. Chem. Phys.}\ }\textbf {\bibinfo {volume} {13}},\
  \bibinfo {pages} {19114} (\bibinfo {year} {2011})}\BibitemShut {NoStop}%
\bibitem [{\citenamefont {Ni}\ \emph {et~al.}(2010)\citenamefont {Ni},
  \citenamefont {Ospelkaus}, \citenamefont {Wang}, \citenamefont
  {Qu{\'e}m{\'e}ner}, \citenamefont {Neyenhuis}, \citenamefont {de~Miranda},
  \citenamefont {Bohn}, \citenamefont {Ye},\ and\ \citenamefont
  {Jin}}]{Ni2010}%
  \BibitemOpen
  \bibfield  {author} {\bibinfo {author} {\bibfnamefont {K.-K.}\ \bibnamefont
  {Ni}}, \bibinfo {author} {\bibfnamefont {S.}~\bibnamefont {Ospelkaus}},
  \bibinfo {author} {\bibfnamefont {D.}~\bibnamefont {Wang}}, \bibinfo {author}
  {\bibfnamefont {G.}~\bibnamefont {Qu{\'e}m{\'e}ner}}, \bibinfo {author}
  {\bibfnamefont {B.}~\bibnamefont {Neyenhuis}}, \bibinfo {author}
  {\bibfnamefont {M.~H.~G.}\ \bibnamefont {de~Miranda}}, \bibinfo {author}
  {\bibfnamefont {J.~L.}\ \bibnamefont {Bohn}}, \bibinfo {author}
  {\bibfnamefont {J.}~\bibnamefont {Ye}}, \ and\ \bibinfo {author}
  {\bibfnamefont {D.~S.}\ \bibnamefont {Jin}},\ }\href {\doibase
  10.1038/nature08953} {\bibfield  {journal} {\bibinfo  {journal} {Nature}\
  }\textbf {\bibinfo {volume} {464}},\ \bibinfo {pages} {1324} (\bibinfo {year}
  {2010})}\BibitemShut {NoStop}%
\bibitem [{\citenamefont {DeMarco}\ \emph {et~al.}(1999)\citenamefont
  {DeMarco}, \citenamefont {Bohn}, \citenamefont {Burke}, \citenamefont
  {Holland},\ and\ \citenamefont {Jin}}]{DeMarco1999}%
  \BibitemOpen
  \bibfield  {author} {\bibinfo {author} {\bibfnamefont {B.}~\bibnamefont
  {DeMarco}}, \bibinfo {author} {\bibfnamefont {J.~L.}\ \bibnamefont {Bohn}},
  \bibinfo {author} {\bibfnamefont {J.~P.}\ \bibnamefont {Burke}}, \bibinfo
  {author} {\bibfnamefont {M.}~\bibnamefont {Holland}}, \ and\ \bibinfo
  {author} {\bibfnamefont {D.~S.}\ \bibnamefont {Jin}},\ }\href {\doibase
  10.1103/PhysRevLett.82.4208} {\bibfield  {journal} {\bibinfo  {journal}
  {Phys. Rev. Lett.}\ }\textbf {\bibinfo {volume} {82}},\ \bibinfo {pages}
  {4208} (\bibinfo {year} {1999})}\BibitemShut {NoStop}%
\bibitem [{\citenamefont {Ma}\ \emph {et~al.}(2003)\citenamefont {Ma},
  \citenamefont {Thomas}, \citenamefont {Foot},\ and\ \citenamefont
  {Cornish}}]{Ma2003}%
  \BibitemOpen
  \bibfield  {author} {\bibinfo {author} {\bibfnamefont {Z.-Y.}\ \bibnamefont
  {Ma}}, \bibinfo {author} {\bibfnamefont {A.~M.}\ \bibnamefont {Thomas}},
  \bibinfo {author} {\bibfnamefont {C.~J.}\ \bibnamefont {Foot}}, \ and\
  \bibinfo {author} {\bibfnamefont {S.~L.}\ \bibnamefont {Cornish}},\ }\href
  {https://iopscience.iop.org/article/10.1088/0953-4075/36/16/313} {\bibfield
  {journal} {\bibinfo  {journal} {J. Phys. B: At. Mol. Opt. Phys.}\ }\textbf
  {\bibinfo {volume} {36}},\ \bibinfo {pages} {3533} (\bibinfo {year}
  {2003})}\BibitemShut {NoStop}%
\bibitem [{\citenamefont {Duda}\ \emph {et~al.}(2021)\citenamefont {Duda},
  \citenamefont {Chen}, \citenamefont {Schindewolf}, \citenamefont {Bause},
  \citenamefont {von Milczewski}, \citenamefont {Schmidt}, \citenamefont
  {Bloch},\ and\ \citenamefont {Luo}}]{Duda2021}%
  \BibitemOpen
  \bibfield  {author} {\bibinfo {author} {\bibfnamefont {M.}~\bibnamefont
  {Duda}}, \bibinfo {author} {\bibfnamefont {X.-Y.}\ \bibnamefont {Chen}},
  \bibinfo {author} {\bibfnamefont {A.}~\bibnamefont {Schindewolf}}, \bibinfo
  {author} {\bibfnamefont {R.}~\bibnamefont {Bause}}, \bibinfo {author}
  {\bibfnamefont {J.}~\bibnamefont {von Milczewski}}, \bibinfo {author}
  {\bibfnamefont {R.}~\bibnamefont {Schmidt}}, \bibinfo {author} {\bibfnamefont
  {I.}~\bibnamefont {Bloch}}, \ and\ \bibinfo {author} {\bibfnamefont {X.-Y.}\
  \bibnamefont {Luo}},\ }\href {http://arxiv.org/abs/2111.04301} {\bibfield
  {journal} {\bibinfo  {journal} {arXiv:2111.04301}\ } (\bibinfo {year}
  {2021})}\BibitemShut {NoStop}%
\bibitem [{\citenamefont {Schindewolf}\ \emph {et~al.}(2022)\citenamefont
  {Schindewolf}, \citenamefont {Bause}, \citenamefont {Chen}, \citenamefont
  {Duda}, \citenamefont {Karman}, \citenamefont {Bloch},\ and\ \citenamefont
  {Luo}}]{Schindewolf2022}%
  \BibitemOpen
  \bibfield  {author} {\bibinfo {author} {\bibfnamefont {A.}~\bibnamefont
  {Schindewolf}}, \bibinfo {author} {\bibfnamefont {R.}~\bibnamefont {Bause}},
  \bibinfo {author} {\bibfnamefont {X.-Y.}\ \bibnamefont {Chen}}, \bibinfo
  {author} {\bibfnamefont {M.}~\bibnamefont {Duda}}, \bibinfo {author}
  {\bibfnamefont {T.}~\bibnamefont {Karman}}, \bibinfo {author} {\bibfnamefont
  {I.}~\bibnamefont {Bloch}}, \ and\ \bibinfo {author} {\bibfnamefont {X.-Y.}\
  \bibnamefont {Luo}},\ }\href {https://arxiv.org/abs/2201.05143} {\bibfield
  {journal} {\bibinfo  {journal} {arXiv:2201.05143}\ } (\bibinfo {year}
  {2022})}\BibitemShut {NoStop}%
\bibitem [{\citenamefont {Zhang}\ \emph {et~al.}(2021)\citenamefont {Zhang},
  \citenamefont {Chen}, \citenamefont {Yao},\ and\ \citenamefont
  {Chin}}]{Zhang2021}%
  \BibitemOpen
  \bibfield  {author} {\bibinfo {author} {\bibfnamefont {Z.}~\bibnamefont
  {Zhang}}, \bibinfo {author} {\bibfnamefont {L.}~\bibnamefont {Chen}},
  \bibinfo {author} {\bibfnamefont {K.-X.}\ \bibnamefont {Yao}}, \ and\
  \bibinfo {author} {\bibfnamefont {C.}~\bibnamefont {Chin}},\ }\href
  {https://www.nature.com/articles/s41586-021-03443-0?proof=tNature} {\bibfield
   {journal} {\bibinfo  {journal} {Nature}\ }\textbf {\bibinfo {volume}
  {592}},\ \bibinfo {pages} {708} (\bibinfo {year} {2021})}\BibitemShut
  {NoStop}%
\bibitem [{\citenamefont {Durastante}\ \emph {et~al.}(2020)\citenamefont
  {Durastante}, \citenamefont {Politi}, \citenamefont {Sohmen}, \citenamefont
  {Ilzh{\"o}fer}, \citenamefont {Mark}, \citenamefont {Norcia},\ and\
  \citenamefont {Ferlaino}}]{Durastante2020}%
  \BibitemOpen
  \bibfield  {author} {\bibinfo {author} {\bibfnamefont {G.}~\bibnamefont
  {Durastante}}, \bibinfo {author} {\bibfnamefont {C.}~\bibnamefont {Politi}},
  \bibinfo {author} {\bibfnamefont {M.}~\bibnamefont {Sohmen}}, \bibinfo
  {author} {\bibfnamefont {P.}~\bibnamefont {Ilzh{\"o}fer}}, \bibinfo {author}
  {\bibfnamefont {M.~J.}\ \bibnamefont {Mark}}, \bibinfo {author}
  {\bibfnamefont {M.~A.}\ \bibnamefont {Norcia}}, \ and\ \bibinfo {author}
  {\bibfnamefont {F.}~\bibnamefont {Ferlaino}},\ }\href {\doibase
  10.1103/PhysRevA.102.033330} {\bibfield  {journal} {\bibinfo  {journal}
  {Phys. Rev. A}\ }\textbf {\bibinfo {volume} {102}},\ \bibinfo {pages}
  {033330} (\bibinfo {year} {2020})}\BibitemShut {NoStop}%
\bibitem [{\citenamefont {Frisch}\ \emph {et~al.}(2015)\citenamefont {Frisch},
  \citenamefont {Mark}, \citenamefont {Aikawa}, \citenamefont {Baier},
  \citenamefont {Grimm}, \citenamefont {Petrov}, \citenamefont {Kotochigova},
  \citenamefont {Qu\'em\'ener}, \citenamefont {Lepers}, \citenamefont
  {Dulieu},\ and\ \citenamefont {Ferlaino}}]{Frisch2015}%
  \BibitemOpen
  \bibfield  {author} {\bibinfo {author} {\bibfnamefont {A.}~\bibnamefont
  {Frisch}}, \bibinfo {author} {\bibfnamefont {M.}~\bibnamefont {Mark}},
  \bibinfo {author} {\bibfnamefont {K.}~\bibnamefont {Aikawa}}, \bibinfo
  {author} {\bibfnamefont {S.}~\bibnamefont {Baier}}, \bibinfo {author}
  {\bibfnamefont {R.}~\bibnamefont {Grimm}}, \bibinfo {author} {\bibfnamefont
  {A.}~\bibnamefont {Petrov}}, \bibinfo {author} {\bibfnamefont
  {S.}~\bibnamefont {Kotochigova}}, \bibinfo {author} {\bibfnamefont
  {G.}~\bibnamefont {Qu\'em\'ener}}, \bibinfo {author} {\bibfnamefont
  {M.}~\bibnamefont {Lepers}}, \bibinfo {author} {\bibfnamefont
  {O.}~\bibnamefont {Dulieu}}, \ and\ \bibinfo {author} {\bibfnamefont
  {F.}~\bibnamefont {Ferlaino}},\ }\href {\doibase
  10.1103/PhysRevLett.115.203201} {\bibfield  {journal} {\bibinfo  {journal}
  {Phys. Rev. Lett.}\ }\textbf {\bibinfo {volume} {115}},\ \bibinfo {pages}
  {203201} (\bibinfo {year} {2015})}\BibitemShut {NoStop}%
\end{thebibliography}%

\end{document}